\documentclass[aps,prl,floatfix,epsfig,twocolumn,showpacs,preprintnumbers]{revtex4}
\usepackage{amssymb}
\usepackage{graphicx}
\usepackage{amsmath}

\begin{document}

\title{CaTe: a new topological node-line and Dirac semimetal}
\author{Yongping Du$^{1}$, Feng Tang$^{1}$, Di Wang$^{1}$, Li Sheng$^{1,2}$,
Er-jun Kan$^{3}$, Chun-Gang Duan$^{4}$, Sergey Y. Savrasov$^{5}$ and
Xiangang Wan$^{1,2}$}
\thanks{Corresponding author: xgwan@nju.edu.cn}
\affiliation{$^{1}$National Laboratory of Solid State Microstructures and Department of
Physics, Nanjing University, Nanjing 210093, China\\
$^{2}$Collaborative Innovation Center of Advanced Microstructures, Nanjing
University, Nanjing 210093, China\\
$^{3}$Key Laboratory of Soft Chemistry and Functional Materials (Ministry of
Education), and Department of Applied Physics, Nanjing University of Science
and Technology, Nanjing, Jiangsu 210094, P. R. China.\\
$^{4}$Key Laboratory of Polar Materials and Devices, Ministry of Education,
East China Normal University, Shanghai 200062, China\\
$^{5}$Department of Physics, University of California, Davis, One Shields
Avenue, Davis, California 95616, USA}

\begin{abstract}
Topological semimetals recently stimulate intense research activities.
Combining first-principles calculations and effective model analysis, we
predict that CaTe is topological node-line semimetal when spin-orbit
coupling (SOC) is ignored. We also obtain the nearly flat surface state
which has the drumhead characteristic. When SOC is included, three node
lines evolve into a pair of Dirac points along the $M-R$ line. These Dirac
points are robust and protected by $C_{4}$ rotation symmetry. Once this
crystal symmetry is broken, the Dirac points will be eliminated, and the
system becomes a strong topological insulator.
\end{abstract}

\date{\today }
\maketitle

\section{Introduction}

Topological insulator (TI) has attracted broad interest in recent years\cite%
{RE-1,RE-2}. The unique property of TI is that the bulk state has a energy
gap while the surface state is gapless. The topological property also have
been proposed for three dimensional (3D) semimetal\cite%
{Weyl-1,Weyl-2,RE-3,Weyl-response,DSM-rev}. Up to now, three kinds of
topological semimetal have been discovered, i.e., 3D Dirac semimetal (DSM)%
\cite{DSM-rev,DSM-1,Na3Bi,Cd3As2,BaAgBi,Cava}, Weyl semimetal (WSM)\cite%
{Weyl-1,Weyl-2,HgCr2Se4,balets} and node-line semimetal (NLS)\cite%
{NSL-1,NSL-2,NSL-3,NSL-4}. The 3D DSM has four-fold degeneracy point formed
by two double degeneracy linear band crossing. Combining the crystal
symmetry and time reversal symmetry, 3D DSM can be robust against external
perturbations. Based on band structural calculation, several materials have
been proposed to be 3D DSM\cite{DSM-rev,DSM-1,Na3Bi,Cd3As2,BaAgBi,Cava} and
some of them already been confirmed by experiments\cite%
{DSMExp-1,DSMExp-3,DSMExp-4,DSMExp-5}. If one breaks time reversal symmetry%
\cite{Weyl-1,HgCr2Se4,balets,axion} or inversion symmetry\cite%
{Weyl-inver-1,Weyl-inver-2,TaAs,TaAs-hasan}, the double degeneracy bands
will split, consequently the 3D DSM evolves into WSM. Very recently, the
predictions about WSM in TaAs family \cite{TaAs,TaAs-hasan} had been
confirmed experimentally \cite{TaAs-exp-1,TaAs-exp-2,TaAs-exp-3,TaAs-exp-4}.

Unlike DSM and WSM whose band crossing points distribute at separate $k$
points in the Brillouin zone (BZ), for the NLS, the crossing points around
the Fermi level form a closed loop. Several compounds had been proposed as
NLS included MTC\cite{NSL-2}, Bernal graphite\cite{NSL-mat-2},
hyperhoneycomb lattices\cite{NSL-mat-3} and antiperovskite Cu$_{3}$PdN\cite%
{NSL-3,NSL-4} and Cu$_{3}$NZn\cite{NSL-4}. When SOC is neglected, for the
system with band inversion, time reversal symmetry together with inversion
symmetry or mirror symmetry will guarantee node line in 3D BZ\cite%
{NSL-2,NSL-3,NSL-4,TaAs,NSL-mat-4,NSL-mat-5}. Same with TI\ and WSM, NLS
also has an characteristic surface state, namely, \emph{drumhead} like state%
\cite{NSL-1,NSL-2,NSL-3,NSL-4}. Such 2D flat band surface state may become a
route to achieve high temperature superconductivity\cite{NSL-app-1,NSL-app-2}%
.

In this article, based on first-principles calculations and effective model
analysis, we propose that CaTe in CsCl-type structure is a NLS with drumhead
like surface flat bands when SOC is ignored. As shown in Fig. 1(b), around
the $M$ point, there are three node-line rings, which is perpendicular to
each other. When SOC is included, these three node-line rings evolve into
two Dirac points along the $M-R$ line. The Dirac points are robust and
protected by the $C_{4}$ rotational symmetry. If the $C_{4}$ symmetry is
broken, the system becomes a strong topological insulator with $Z_{2}$
indices (1;000).

\begin{figure}[tbh]
\center\includegraphics[scale=0.4]{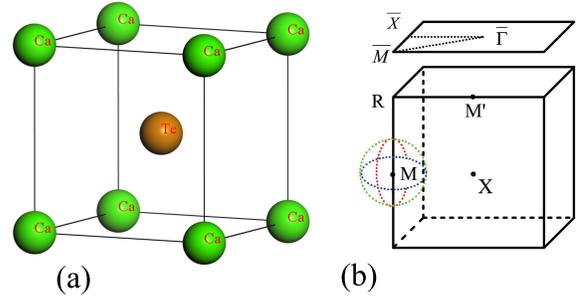}
\caption{(color online). (a) Crystal structure of CaTe in CsCl type phase. (b)The 3D BZ and
projected (001) two dimensional (2D) BZ of CaTe. Three dash circles are the
scheme of the three line nodes around the $M$ point. The blue circle is
parallel to $k_{z}=0$ plane, red circle is parallel to $k_{x}=0$ plane and
the green circle is parallel to $k_{y}=0$ plane.}
\label{Fig1-cry}
\end{figure}

\section{Crystal structure and method}

As one member of the alkaline-earth chalcogenides, CaTe have attracted
tremendous interests because of its technological applications ranging from
catalysis to luminescence\cite%
{Crystal-1,Crystal-2,Crystal-3,Crystal-4,Crystal-5}. CaTe undergoes a phase
transition from NaCl-type structure at ambient conditions to CsCl-type
structure at hydrostatic pressure about 33 GPa\cite{Crystal-1,Crystal-2}.
The structure of CaTe in CsCl-type is shown in Fig.1(a). The space group of
this phase is $Pm\overline{3}m$ (NO. 221). The electronic band structure
calculations have been carried out using the full potential linearized
augmented plane wave method as implemented in WIEN2K package \cite{WIEN2K}.
To obtain accurate band inversion strength and band order, the modified
Becke-Johnson exchange potential together with local-density approximation
for the correlation potential (MBJLDA) has been applied \cite{mBJ}. The
plane-wave cutoff parameter R$_{MT}$K$_{max}$ is set to 7 and a $16\times
16\times 16$ mesh was used for the BZ integral. The SOC interaction is
included by using the second-order variational procedure.

\section{Electronic structure}

Firstly, we calculate the band structure of CaTe and show the result without
SOC in Fig. 2(a). By checking the wave functions, we find that the valence
bands and conduction bands are mainly contributed by 5$p_{z}$ (blue) state
of Te and 3$d_{z^{2}}$ (red) state of Ca, respectively, as shown in Fig.
2(a). The band inversion happened at $M$ point where the energy of Te-5$%
p_{z} $ state is higher than the energy of Ca-3$d_{z^{2}}$ state by about
0.75 eV. Interestingly, this kind of band inversion is not caused by the
SOC, which is different from most topological materials\cite{RE-1,RE-2}. We
calculate the electronic structure of CaTe by applying tensile strain to
check the origin of band inversion at $M$ point. The energy difference
between Te-5$p_{z}$ state and Ca-3$d_{z^{2}}$ state decreases as increasing
the tensile strain. We find that when a$\geq $1.05a$_{0}$, the band
inversion at M point disappear. With the time reversal symmetry and
inversion symmetry, the band inversion results in node lines as proved in
Ref. \cite{NSL-2}.

\begin{figure}[tbh]
\center\includegraphics[scale=0.4]{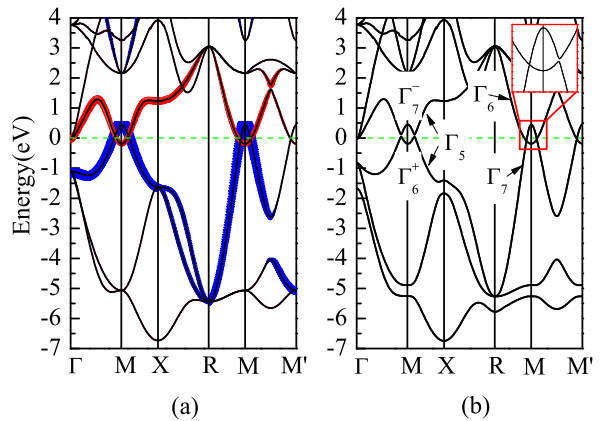}
\caption{(color online). (a) Electronic structure of CaTe without SOC. The weights of Te-5$%
p_{z}$ (Ca-3$d_{z^{2}}$)state is proportional to the width of blue (red)
curves. (b) Electronic structure of CaTe with SOC. Along $\Gamma -M$, $M-X$
and $M-M^{\prime }$, a small gap is opened. The Dirac point at the $M-R$
line is protected by $C_{4}$ rotational symmetry. The inset is the detailed
structure around the M point. (see main text for detailed description).}
\label{Fig2-band}
\end{figure}

The effective Hamiltonian of node line around $M$ point can be established
by using the $\overset{\rightharpoonup }{k}\cdot $ $\overset{\rightharpoonup
}{p}$ method. Considering the crystal symmetry at $M$ point and time
reversal symmetry, the effective Hamiltonian can be written as following
form:%
\begin{equation*}
H(\overset{\rightharpoonup }{k})=g_{0}(\overset{\rightharpoonup }{k})\sigma
_{0}+g_{x}(\overset{\rightharpoonup }{k})\sigma _{x}+g_{z}(\overset{%
\rightharpoonup }{k})\sigma _{z},
\end{equation*}%
where the $\sigma _{x}\ $and $\sigma _{z}$ are Pauli matrices, $\sigma _{0}$
is unit matrix. $g_{0}(\overset{\rightharpoonup }{k}%
)=M_{0}-B_{0}(k_{x}^{2}+k_{y}^{2})-C_{0}k_{z}^{2}$, $g_{x}(\overset{%
\rightharpoonup }{k})=Ak_{x}k_{y}k_{z}$, $g_{z}(\overset{\rightharpoonup }{k}%
)=M_{z}-B_{z}(k_{x}^{2}+k_{y}^{2})-C_{z}k_{z}^{2}$. This system has both of
time reversal symmetry and inversion symmetry, thus, the component of $%
\sigma _{y}$ is zero\cite{NSL-2}. The eigenvalues of this $2\times 2$
effective Hamiltonian are $E(\overset{\rightharpoonup }{k})=g_{0}(\overset{%
\rightharpoonup }{k})\pm \sqrt{g_{x}^{2}(\overset{\rightharpoonup }{k}%
)+g_{z}^{2}(\overset{\rightharpoonup }{k})}$. When $g_{x}(\overset{%
\rightharpoonup }{k})=0$ and $g_{z}(\overset{\rightharpoonup }{k})=0$, the
nodal line will emergent. It can be easily checked that the equation $g_{z}(%
\overset{\rightharpoonup }{k})=0$ has solution only when $M_{z}B_{z}>0$ and $%
M_{z}C_{z}>0$. And $M_{z}B_{z}>0$ and $M_{z}C_{z}>0$ are also the condition
of band inversion. On the other hand, $g_{x}(\overset{\rightharpoonup }{k}%
)=0 $ confine the node lines in three mutually perpendicular planes (namely,
$k_{x}$=0 plane, $k_{y}$=0 plane and $k_{z}$=0 plane ) as illustrated in
Fig. 1(b). Due to the fact that $g_{0}(\overset{\rightharpoonup }{k})$ does
not equal to zero which breaks the electron-hole symmetry, consequently, the
node lines have finite energy dispersion.

When SOC is considered, three node lines evolve into two Dirac points at $%
M-R $ line as shown in Fig. 2(b). At $M$ point, the two states near Fermi
level belong to irreducible representation $\Gamma _{7}^{-}$ and $\Gamma
_{6}^{+}$, respectively. While along the $M-X$ line, two bands have the same
irreducible representation $\Gamma _{5}\ $as shown in Fig. 2(b), thus they
can hybridize with each other and open a small gap (about 50meV). For both $%
\Gamma -M$ line and $M-M^{\prime }$ line, the two bands around Fermi level
are also belong to the same irreducible representation, thus there is not
band crossing along $\Gamma -M$ line and $M-M^{\prime }$ line. Since the
band splitting is determined by the SOC, thus one can achieve the NLS by
doping the lighter atoms such as Se, S.

Along the $M-R$ line, which reserve the $C_{4}$ rotation symmetry, two
states with $\Gamma _{7}^{-}$ and $\Gamma _{6}^{+}$ at $M$ point evolve into
$\Gamma _{7}$ and $\Gamma _{6}$, thus the hybridization between these two
bands is forbidden, there is a Dirac point as shown in Fig.2(b). When the $%
C_{4}$ rotational symmetry is broken, like by strain effect, the band
crossing point will disappear, and this 3D DSM will become a strong TI with
topological indices $Z_{2}$ to be(1;0,0,0).

To understand the band inversion at $M$ point and the topological property
of this system, we derive a low energy effective Hamiltonian at $M$ point
based on the projection-operator method\cite{BaAgBi}. $M$ point has $D_{4h}$
symmetry and also time reversal symmetry. As discussed above, at $M$ point, $%
\Gamma _{7}^{-}$ symmetry state has angular momentum $j_{z}=\pm 3/2$ and $%
\Gamma _{6}^{+}$ symmetry state has angular momentum $j_{z}=\pm 1/2$.
Therefore using the basis of ($|j_{z}=-\frac{1}{2}\rangle _{d},|j_{z}=+\frac{%
1}{2}\rangle _{d},|j_{z}=-\frac{3}{2}\rangle _{p},|j_{z}=+\frac{3}{2}\rangle
_{p}$, the effective Hamiltonian around $M$ point can be written as (see
APPENDIX for detail.):
\begin{widetext}
\begin{equation*}
H_{eff}=\left(
\begin{matrix}
M_{1}(\overset{\rightharpoonup }{k}) & 0 & Ak_{+}+B(\overset{\rightharpoonup }{k}) & D(\overset{\rightharpoonup }{k}) \\
0 & M_{1}(\overset{\rightharpoonup }{k}) & D^{\ast }(\overset{\rightharpoonup }{k}) & -Ak_{-}-B^{\ast }(\overset{\rightharpoonup }{k})
\\
Ak_{-}+B^{\ast }(\overset{\rightharpoonup }{k}) & D(\overset{\rightharpoonup }{k}) & M_{2}(\overset{\rightharpoonup }{k}) & 0 \\
D^{\ast }(\overset{\rightharpoonup }{k}) & -Ak_{+}-B(\overset{\rightharpoonup }{k}) & 0 & M_{2}(\overset{\rightharpoonup }{k}),%
\end{matrix}%
\right)
\end{equation*}%
\end{widetext}where $M_{1}(\overset{\rightharpoonup }{k}%
)=M_{10}+M_{11}(k_{x}^{2}+k_{y}^{2})+M_{12}k_{z}^{2}$, $M_{2}(\overset{%
\rightharpoonup }{k})=M_{20}+M_{21}(k_{x}^{2}+k_{y}^{2})+M_{22}k_{z}^{2}$, $%
B(\overset{\rightharpoonup }{k}%
)=B_{1}k_{+}k_{z}^{2}+B_{2}(k_{x}^{3}+ik_{y}^{3})+iB_{3}k_{-}k_{x}k_{y}$, $D(%
\overset{\rightharpoonup }{k}%
)=D_{1}(k_{x}^{2}-k_{y}^{2})k_{z}+iD_{2}k_{x}k_{y}k_{z}$, $k_{\pm }=k_{x}\pm
ik_{y}$. Along the $k_{z}$ axis (where $k_{x}$=0, $k_{y}$=0) the effective
Hamiltonian is diagonal, and the eigenvalues are $E(\overset{\rightharpoonup
}{k})=M_{1}(\overset{\rightharpoonup }{k})$ and $E(\overset{\rightharpoonup }%
{k})=M_{2}(\overset{\rightharpoonup }{k})$. As mentioned above, the Dirac
point is on the $M-R$ line, thus it is interesting to discuss the effective
model along this line. Since there is the band inversion between $\left\vert
j_{z}=\pm \frac{1}{2}\right\rangle _{d}$ and $\left\vert j_{z}=\pm \frac{3}{2%
}\right\rangle _{p}$ at $M$ point, it is easy to obtain that $M_{10}<M_{20}$%
, $M_{22}<0<M_{12}$, and the Dirac points locate at $\overset{%
\rightharpoonup }{k_{c}}=(\frac{\pi }{a},\frac{\pi }{a},k_{zc}$=$\pm \sqrt{%
\frac{M_{20}-M_{10}}{M_{12}-M_{22}}})$. Neglecting the high-order terms, $E(%
\overset{\rightharpoonup }{k_{c}}+\overset{\rightharpoonup }{\delta k})$ can
be expressed as $(M_{12}+M_{22})k_{zc}\delta k_{z}\pm \sqrt{%
(M_{12}-M_{22})^{2}k_{zc}^{2}\delta k_{z}^{2}+A^{2}(\delta k_{x}^{2}+\delta
k_{y}^{2})}$, where $\delta k_{x,y,z}$ are small displacement from $\overset{%
\rightharpoonup }{k_{c}}$. In the vicinity of $\overset{\rightharpoonup }{%
k_{c}}$, the band dispersion is a linear, thus our effective Hamiltonian is
nothing but 3D\ massless Dirac fermions.

\begin{figure}[tbph]
\centering\includegraphics[scale=0.4]{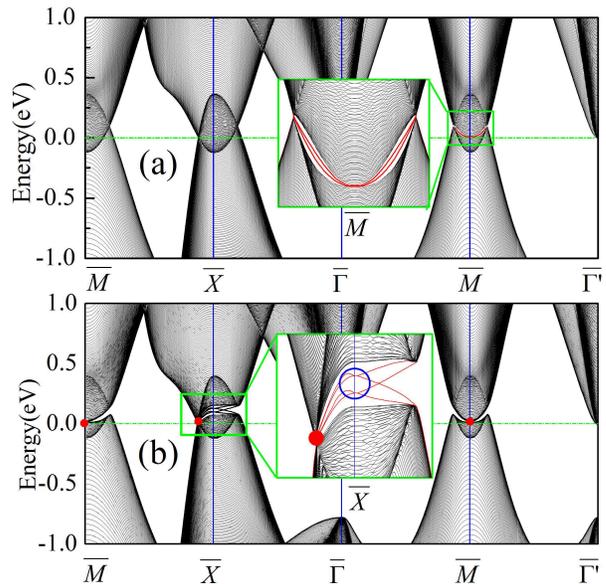}
\caption{(color online). The surface state of (001) surface of CaTe. (a) The surface state
is calculated without SOC. The flat surface state around Fermi level is
denoted by red color. The inset is the detailed band structure around the $%
\overline{M}$ point. (b) The surface state is calculated with SOC. The red
dots are the projected bulk Dirac nodes. The red lines between the bulk gap
at $\overline{X}$ point are surface states and the blue circle denotes the
surface Dirac cones. The inset is the detailed band structure around the $%
\overline{X}$ point.}
\label{Fig3-surface-state}
\end{figure}

The band inversion at $M$ point and the Dirac nodes in CaTe suggest the
existence of topological nontrivial surface state. To study the surface
states in CaTe we use a 200-unit-cells-thick (001) slab with top (bottom)
surface terminated by Ca (Te) atoms. The surface state is then calculated by
using the tight-binding method. The hopping parameters are determined from a
maximally localized Wannier functions (MLWFs)\cite{MLWFs}, which are
projected from the Bloch state derived from first-principles calculations.

Fig. 3(a)/(b) shows the surface state of CaTe (001) surface without/with
SOC, respectively. When SOC is ignored, the system is a NLS, and possess
nearly flat surface band around the Fermi energy. As shown in Fig. 3(a), our
numerical results find that the nearly flat surface "drumhead" state appears
in the interiors of the projected nodal line rings on the (001) surface
around the $\overline{M}$ point. Since the slab we used has two surfaces,
there are two surface states as shown in the red lines in the Fig. 3(a). The
particle-hole symmetry is broken by nonzero term $g_{0}(\overset{%
\rightharpoonup }{k})$, thus these two surface bands are not perfect flat
with about 70 meV bandwidth. This type of 2D flat bands are proposed as a
novel route to achieve high temperature superconductivity\cite%
{NSL-app-1,NSL-app-2}.

When the SOC is included, three node lines are gapped out and become a pair
of Dirac points along the $M-R$ line, thus the NLS become a 3D DSM. There is
bulk Dirac node projected on $\overline{M}$ point (the red dot) as shown in
Fig. 3(b). Along the $\overline{M}-\overline{X}$, there is also a projected
bulk Dirac node, which locate near the $\overline{X}$ point denoted by red
dots. Fig. 3(b) is clearly shows that the gapped bulk states along the $%
\overline{\Gamma }-\overline{X}$ direction and the existence of surface
Dirac cones (in the blue circle) due to the topologically nontrivial $Z_{2}$
indices, like the same case in Na$_{3}$Bi\cite{Na3Bi} and Cd$_{3}$As$_{2}$%
\cite{Cd3As2}.

\section{Conclusion}

In summary, based on first-principles calculations and effective model
analysis, we suggest that CaTe in CsCl-type structure is a NLS when SOC is
ignored. There are three node-line rings which are perpendicular to each
other around the $M$ point. With band inversion at $M$ point, this NLS is
protected by the time reversal symmetry and inversion symmetry. When the SOC
is included, three node-line rings become a pair of Dirac points. These
Dirac nodes are robust and protected by the $C_{4}$ crystal symmetry and the
system become a DSM.

\section{Acknowledgement}

The work was supported by the National Key Project for Basic Research of
China (Grant No. 2014CB921104), NSFC under Grants No. 11525417 and 11374147.
The project is also funded by Priority Academic Program Development of
Jiangsu Higher Education Institutions. S.S. were supported by NSF DMR (Grant
No. 1411336). Y.D. is supported by the program B for Outstanding PhD
candidate of Nanjing University.

\section{APPENDIX}

The conduction and valence bands of CaTe at $M$ point are mainly contributed
by four states: $|j_{z}=-\frac{1}{2}\rangle _{d}$, $|j_{z}=+\frac{1}{2}%
\rangle _{d}$, $|j_{z}=-\frac{3}{2}\rangle _{p}$ and $|j_{z}=+\frac{3}{2}%
\rangle _{p}$, we thus use these states as the basis to build the effective
model Hamiltonian at the $M$\ point of BZ. As a $4\times 4$ Hermitian
matrix, the effective Hamiltonian can be expended as $H=\underset{i=1}{%
\overset{16}{\sum }}f_{i}(\overset{\rightharpoonup }{k})\Gamma _{i}$, where $%
f_{i}(\overset{\rightharpoonup }{k})$ are function of momentum \textbf{k}. $%
\Gamma _{i}$ are Dirac matrices, which can be written as the direct product
of $\sigma _{i}$ and $\tau _{j}$ ($\sigma _{i=1,2,3,4}$, $\tau _{j=1,2,3,4}$
are unit matrix $\sigma _{0}$ and three Pauli matrices $\sigma _{x}$, $%
\sigma _{y}$ and $\sigma _{z}$).

Under the operation of crystal symmetry and time reversal symmetry, the
Hamiltonian should be invariant. This requires the function $f_{i}(\overset{%
\rightharpoonup }{k})$ and the associated $\Gamma _{i}$ matrices belong to
the same irreducible representation. Thus the key problem is to determine
the irreducible representation for $f_{i}(\overset{\rightharpoonup }{k})$
and $\Gamma _{i}$ matrices, which can be done by the projection-operator
method.

\begin{table*}[tbh]
\centering%
\begin{tabular}{ccc}
\hline
$\Gamma $ matrices & representation & T \\ \hline
$\Gamma _{1}$,$\Gamma _{13}$ & $R_{1}$ & + \\
$\Gamma _{4},\Gamma _{16}$ & $R_{2}$ & - \\
$\{\Gamma _{2},\Gamma _{3}\},\{\Gamma _{14},\Gamma _{15}\}$ & $R_{5}$ & - \\
$\Gamma _{7}$ & $R_{10}$ & - \\
$\Gamma _{11}$ & $R_{10}$ & + \\
$\Gamma _{6}$ & $R_{11}$ & - \\
$\Gamma _{10}$ & $R_{11}$ & + \\
$\{\Gamma _{8},\Gamma _{9}\}$ & $R_{12}$ & - \\
$\{\Gamma _{5},\Gamma _{12}\}$ & $R_{12}$ & + \\ \hline
$d(k)$ & representation & T \\ \hline
$C$, $k_{z}^{2}$, $k_{x}^{2}+k_{y}^{2}$ & $R_{1}$ & + \\
$k_{x}^{2}-k_{y}^{2}$ & $R_{3}$ & + \\
$k_{x}k_{y}$ & $R_{4}$ & + \\
$\{k_{x}k_{z}$, $k_{y}k_{z}\}$ & $R_{5}$ & + \\
$k_{z},k_{z}^{3},(k_{x}^{2}+k_{y}^{2})k_{z}$ & $R_{9}$ & - \\
$k_{x}k_{y}k_{z}$ & $R_{10}$ & - \\
$\{k_{x}^{2}-k_{y}^{2}\}k_{z}$ & $R_{11}$ & - \\
$%
(k_{x},k_{y}),(k_{x}^{3},k_{y}^{3}),(k_{x}^{2}k_{y},k_{y}^{2}k_{x}),(k_{x}k_{z}^{2},k_{y}k_{z}^{2})
$ & $R_{12}$ & - \\ \hline
\end{tabular}%
\caption{The character table of Dirac $\Gamma $ \ matrices and the
polynomials of the momentum $k$ for CaTe at M point.}
\end{table*}

Because the SOC is included, we use the double space group. Under the
projection-operator method, we present the irreducible representation of
Dirac $\Gamma $\ matrices and polynomials of $\overset{\rightharpoonup }{k}$%
, and their transformation under time reversal in Table I. With the Table I,
the effective model Hamiltonian of CaTe at $M$ point can be easily expressed
as: $H=f_{1}(\overset{\rightharpoonup }{k})\Gamma _{1}+f_{13}(\overset{%
\rightharpoonup }{k})\Gamma _{13}+f_{8}(\overset{\rightharpoonup }{k})\Gamma
_{8}-f_{9}(\overset{\rightharpoonup }{k})\Gamma
_{9}+D_{1}(k_{x}^{2}-k_{y}^{2})k_{z}\Gamma _{6}-D_{2}k_{x}k_{y}k_{z}\Gamma
_{7}$, where $f_{1}(\overset{\rightharpoonup }{k}%
)=C_{1}+m_{1}(k_{x}^{2}+k_{y}^{2})+n_{1}k_{z}^{2}$, $f_{13}(\overset{%
\rightharpoonup }{k})=C_{13}+m_{13}(k_{x}^{2}+k_{y}^{2})+n_{13}k_{z}^{2}$, $%
f_{8}(\overset{\rightharpoonup }{k}%
)=Ak_{x}+B_{1}k_{x}^{3}+B_{2}k_{x}k_{z}^{2}+B_{3}k_{y}^{2}k_{x}$, $f_{9}(%
\overset{\rightharpoonup }{k}%
)=Ak_{y}+B_{1}k_{y}^{3}+B_{2}k_{y}k_{z}^{2}+B_{3}k_{x}^{2}k_{y}$. $\Gamma
_{1}=\sigma _{0}\otimes \tau _{0}$, $\Gamma _{13}=\sigma _{3}\otimes \tau
_{0}$, $\Gamma _{8}=\sigma _{1}\otimes \tau _{3}$, $\Gamma _{9}=\sigma
_{2}\otimes \tau _{0}$, $\Gamma _{6}=\sigma _{1}\otimes \tau _{1}$, $\Gamma
_{7}=\sigma _{1}\otimes \tau _{2}$. Compare with the effective Hamiltonian,
we have $M_{10}(M_{20})=C_{1}\pm C_{13}$, $M_{11}(M_{21})=m_{1}\pm m_{13}$, $%
M_{12}(M_{22})=n_{1}\pm n_{13}$.

\end{document}